\documentclass[twoside]{ilcws07}
\usepackage[latin1]{inputenc}
\usepackage[dvips]{graphicx,epsfig,color}
\usepackage{wrapfig,rotating}
\usepackage{amssymb,amsmath,array}

\newcommand{\bn}{{\bar n}}

\begin{document}

\thispagestyle{empty}
\setcounter{page}{0}
\def\thefootnote{\fnsymbol{footnote}}

\begin{flushright}
MPP-2007-151\\
MIT-CTP 3866\\
%arXiv:yymm.nnnn [hep-ph]
\end{flushright}

\vspace{1cm}

\begin{center}

{\large {\bf Factorization Approach for Top Mass Reconstruction
\\[2mm] 
at High Energies
}}
\footnote{talk given by A.~H.~Hoang at the {\em LCWS07}, 
May 2007, DESY, Hamburg, Germany}

\vspace{1cm}

{\sc 

Sean Fleming$^{1\,}$%
\footnote{
email: fleming@physics.arizona.edu}
, Andre H. Hoang$^{2\,}$%
\footnote{
email: ahoang@mppmu.mpg.de}
, Sonny Mantry$^{1\,}$%
\footnote{
email: mantry@theory.caltech.edu}
, Iain W.~Stewart$^{3\,}$%
\footnote{
email: iains@mit.edu}

}

\vspace*{1cm}

{\it
$^1$Department of Physics, University of Arizona, Tucson, AZ 85721, USA \\[.3em]
$^2$Max-Planck-Institut f\"ur Physik, % (Werner-Heisenberg-Institut),\\ 
F\"ohringer Ring 6, D--80805 Munich, Germany\\[.3em]
$^3$California Institute of Technology, Pasadena, CA 91125, USA\\[.3em]
$^4$Department of Physics, Massachusetts Institute of Technology, 
Boston, MA 02139, USA\\[.3em]
}
\end{center}

\vspace*{0.2cm}

\begin{center} {\bf Abstract} \end{center}
Using effective theories for jets and heavy quarks it is possible to
prove that the double differential top-antitop invariant mass
distribution for the process $e^+e^-\to t\bar t$ in the resonance region for
c.m.\,energies $Q$ much larger than the top mass can 
factorized into perturbatively computable hard 
coefficients and jet functions and a non-perturbative soft function.
For invariant mass prescriptions based on hemispheres defined with
respect to the thrust axis the soft function can be extracted from
massless jet event shape distributions. This approach 
allows in principle for top mass determinations without uncertainties from
hadronization using the reconstruction method and to quantify
the top mass scheme dependence of the measured top quark mass value.

\def\thefootnote{\arabic{footnote}}
\setcounter{footnote}{0}

\newpage

%%%%%%%%%%%%%%%%%%%%%%%%%%%%%%%%%%%%%%%%%%%%%%%%%%%%%%%%%%%%%%%%%%%%%%%

\title{
%%%%   Paper title goes here  %%%%%%%%%%%%%%
Factorization Approach for Top Mass Reconstruction at High Energies} %% 
%***********************************************************************
% AUTHORS INFORMATION AREA
%***********************************************************************
\author{Sean Fleming$^1$, Andre H. Hoang $^2$, Sonny Mantry$^3$, Iain W.~Stewart$^4$
% Optional short acknowledgment: remove next line if non-needed
%\thanks{This work was supported in part by the EU network contract
%  MRTN-CT-2006-035482 (FLAVIAnet).}
% DO NOT MODIFY THE FOLLOWING '\vspace' ARGUMENT
\vspace{.3cm}\\
% Addresses and institutions (remove "1- " in case of a single institution)
1- Department of Physics, University of Arizona, Tucson, AZ 85721, USA \\
2- Max-Planck-Institut f\"ur Physik, (Werner-Heisenberg-Institut) \\
  F\"ohringer Ring 6, M\"unchen, 
  Germany, 80805 %\footnote{Electronic address:  ahoang@mppmu.mpg.de}
%% Remove the next three lines in case of a single institution
\vspace{.1cm}\\
3- California Institute of Technology, Pasadena, CA 91125, USA
\vspace{.1cm}\\
4- Department of Physics, Massachusetts Institute of Technology, 
Boston, MA 02139, USA
}
%%***********************************************************************
% END OF AUTHORS INFORMATION AREA
%***********************************************************************

\maketitle

\begin{abstract}
Using effective theories for jets and heavy quarks it is possible to
prove that the double differential top-antitop invariant mass
distribution for the process $e^+e^-\to t\bar t$ in the resonance region for
c.m.\,energies $Q$ much larger than the top mass can 
factorized into perturbatively computable hard 
coefficients and jet functions and a non-perturbative soft function.
For invariant mass prescriptions based on hemispheres defined with
respect to the thrust axis the soft function can be extracted from
massless jet event shape distributions. This approach 
allows in principle for top mass determinations without uncertainties from
hadronization using the reconstruction method and to quantify
the top mass scheme dependence of the measured top quark mass value.
\end{abstract}

\section{Introduction}
\label{introduction}

Precise measurements of the top quark mass are among the most
important (standard) tasks of the ILC project as the top quark mass
affects a number of interesting observables either directly or
indirectly through quantum effects. To be useful such top mass
measurements have to have small uncertainties, but also need to
provide information to which mass scheme the measured number refers
to. Both aims can be achieved from a threshold scan of the cross
section $\sigma(e^+e^-\to t\bar t)$ for $\sqrt{s}\approx 2m_t$,
from which one expects measurements of the threshold masses, such as
the 1S mass, with uncertainties of about
$100$~MeV~\cite{Hoang:2000yr,Kiyo,AHH,Gounaris,Boogert}. 
Another method is based on mass
reconstruction which relies on the idea that the peak of the invariant
mass distribution of the top decay products is related to the top
quark mass. This method can be applied at any c.m.\,energy and might
also yield uncertainties well below
$1$~GeV~\cite{Chekanov:2002sa}. However, until recently it was
unknown for which mass scheme such measurements can be carried out
with small theoretical uncertainties. This is
because the naive relation between the observable peak of the
invariant mass distribution and the perturbative top quark propagator
pole is affected by hard (i.e. computable) as well as soft
(i.e. non-perturbative) QCD effects and the present MC tools do not
contain the required information in a systematic form.
Obviously the top mass measurements at the
LHC~\cite{Borjanovic:2004ce} suffer from the same problem, but the
associated theoretical systematic uncertainty might be considerably
larger than at the ILC.

\section{Factorization Theorem}
\label{sectionfactheo}
\begin{figure}
  \centerline{ 
   \hspace{1.2cm}\includegraphics[width=8cm]{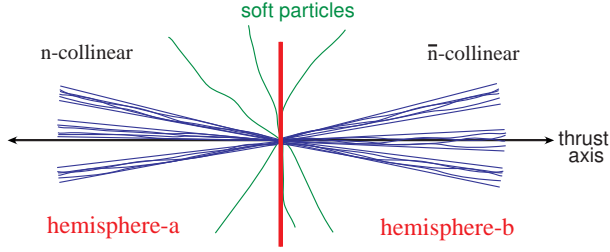}  
  } 
\caption{ Six jet event initiated by a top quark pair, $t\bar t\to bW \bar b
  W\to b qq' \bar b qq'$.  The plane separating the two hemispheres is
  perpendicular to the thrust axis and intersects the thrust axis at
  the interaction point. The total invariant mass inside each
  hemisphere is measured. Our analysis applies equally well to the
  lepton+jets and the dilepton channels (not shown).}
\label{fig:6topjet}
\end{figure}

\begin{figure}
  \centerline{ 
   \includegraphics[width=8cm]{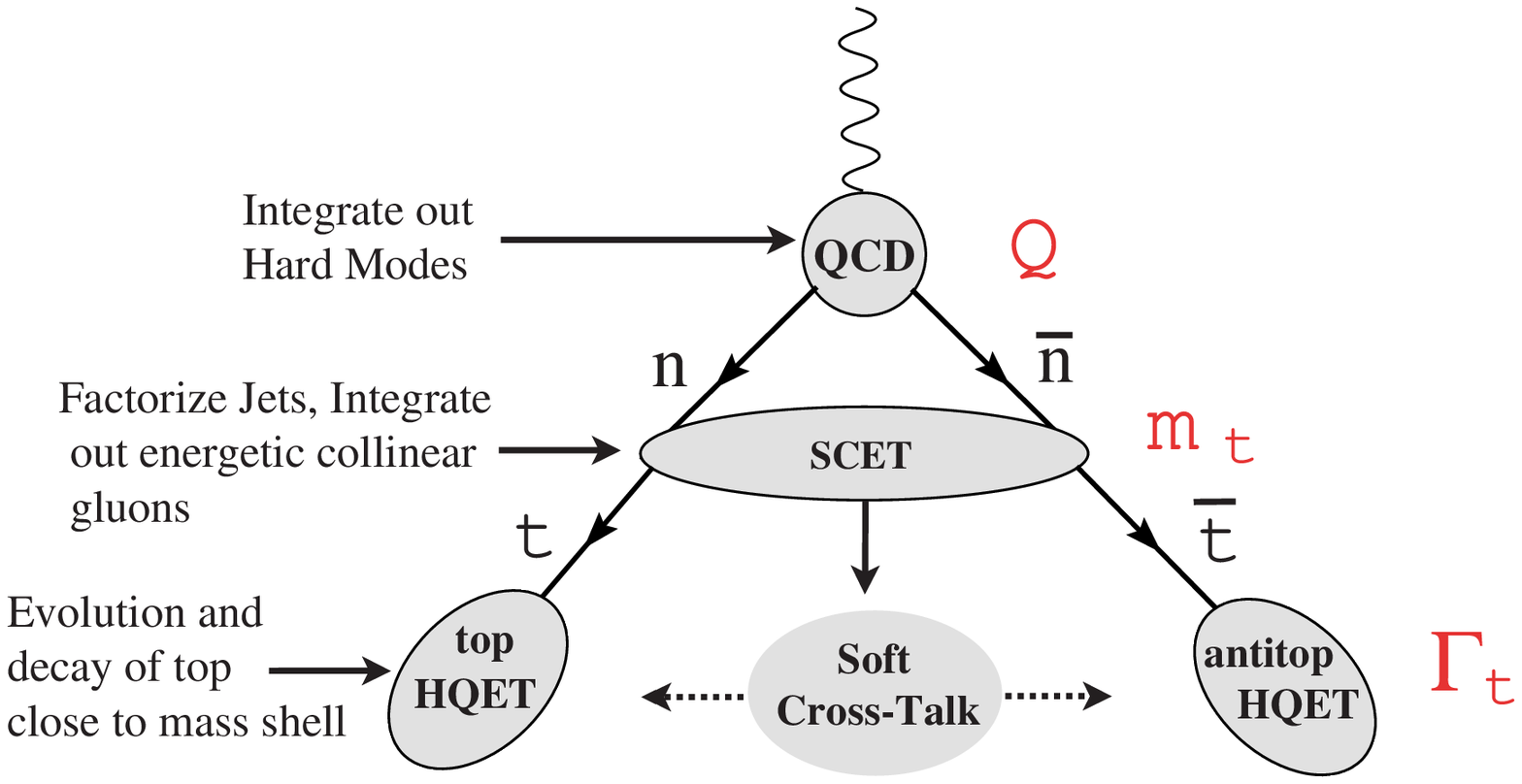}
  } 
\caption{Sequence of effective field theories used to compute the invariant
mass distribution. }
\label{fig:theoryI}
\end{figure}

In Ref.~\cite{Fleming:2007qr} a theoretical formalism was presented
which remedies this situation, as a first step,
for the Linear Collider framework, where one does not need to account
for QCD radiation arising from the initial state. Assuming a
c.m.\,energy $Q\gg m_t$, $m_t$ being the top quark mass, one can
employ the hierarchy of scales 
\begin{equation} 
\label{scales}
Q\,\gg\,m_t\,\gg\,\Gamma_t\, > \,\Lambda_{\rm QCD}
\end{equation}
to establish a factorization theorem for the doubly differential
top-antitop invariant mass distribution in the peak region around the
top resonance:
\begin{equation}
\label{obs1}
\frac{d^2\sigma}{dM^2_t\>dM^2_{\bar{t}}} \,,
  \qquad\qquad M_{t,\bar{t}}^2 - m^2  \sim m\, \Gamma \ll m^2 \,.
\end{equation}
The invariant masses $M_t^2 = ( \sum_{i\in X_t} p_i^\mu)^2$,
$M_{\bar t}^2 =( \sum_{i\in X_{\bar t}} p_i^\mu)^2 $ depend on a
prescription $X_{t,\bar t}$ which associates final state momenta
$p_i^\mu$ to top and antitop invariant masses, respectively. For
invariant masses in the resonance region the events are characterized
by energy deposits predominantly contained in two back-to-back regions
with opening angles $m_t/Q$ associated with the energetic jets or
leptons from the top decay plus collinear radiation, and by additional
soft radiation populating the regions between the jets, see
Fig.~\ref{fig:6topjet}. 
We assume that the prescriptions $X_{t,\bar t}$ assign all soft
radiation to either $M_t^2$ or $M_{\bar t}^2$  where the probability
of radiation being assigned to $X_t$ or $X_{\bar t}$ increases to
unity when it approaches the top or antitop direction. 
The result for the double differential cross-section in the peak
region at all orders in $\alpha_s$ and to leading order in the power
expansion in $m_t\alpha_s/Q$, $m_t^2/Q^2$, $\Gamma_t/m_t$ and
$M_{t,\bar t}-m_t$ is given by~\cite{Fleming:2007qr}
\begin{eqnarray} \label{FactThm}
  \frac{d\sigma}{ dM^2_t\, dM^2_{\bar t}} &=& 
  \sigma_0 \: H_Q(Q,\mu_m) 
  H_m\Big(m_J,\frac{Q}{m_J},\mu_m,\mu\Big)\!
\hspace{2.5cm} \left[\hat s_{t,\bar t} = \frac{M_t^2-m_J^2}{m_J}\right]
  \nonumber\\
 &\times& \int\! d\ell^+ d\ell^- 
   B_+\Big(\hat s_t- \frac{Q\ell^+}{m_J},\Gamma_t,\mu\Big)\:
   B_-\Big(\hat s_{\bar t}-\frac{Q\ell^-}{m_J},\Gamma_t,\mu\Big)  
   S(\ell^+,\ell^-,\mu)\,.
\end{eqnarray}  
In Eq.~(\ref{FactThm}) the normalization factor $\sigma _0$ is the total Born-level
cross-section, the $H_Q$ and $H_m$ are perturbative coefficients describing hard
effects at the scales $Q$ and $m_J$, $B_\pm$ are perturbative jet functions that
describe the evolution and decay of the the top and antitop close to the mass
shell, and $S$ is a nonperturbative soft function describing the soft radiation
between the jets. The result was derived using the hierarchy of
scales~(\ref{scales}), matching QCD onto (Soft-Collinear Effective
Theory) SCET~\cite{Bauer:2000yr} at the scale $\mu=Q$, which
in turn is matched onto (Heavy Quark Effective Theory)
HQET~\cite{Manohar:2000dt} at a scale $\mu_m$ of order $m_t$ generalized for
unstable particle effects associated to the large top width
$\Gamma_t$~\cite{Fadin:1988fn}. An illustration of this scheme is
shown in Fig.~\ref{fig:theoryI}. For details on the (admittedly
non-trivial) derivation and on technical aspects of the factorization
theorem we refer to Ref.~\cite{Fleming:2007qr}. In the
following we will discuss the important ingredients of the
factorization theorem and their physical interpretation and show what
we can learn from them concerning the measurements of the top quark
mass from the reconstruction method. 

\section{Jet Functions and Short-Distance Top Mass}

The coefficients $H_Q$ and $H_m$ in Eq.~(\ref{FactThm}) arise from
matching and running in SCET and HQET down to the low energy scale
$\mu$ where one evaluates the jet functions $B_\pm$ and
the soft function $S$. These hard coefficients only affect the overall 
normalization of the invariant mass distribution and we will therefore
not talk about them here. So let us concentrate on the jet and the
soft functions, which determine the shape of the distribution and the
location of the resonance peak. The jet functions describe the
perturbative contributions of the shape of the invariant mass
distribution and are defined by the  imaginary part of a T-product
vacuum matrix element. For the top quark it is 
\begin{eqnarray}
\label{hqetjet2}
B_+(\hat s_t,\Gamma_t,\mu) & = &
\mbox{Im}\,\left[\,
\frac{- i}{4\pi N_c m_J}  \int\!\! d^4 x \, e^{i r \cdot x} \,
\big\langle 0 \big|T\{\bar{h}_{v_+}(0) W_n(0) W_n^{\dagger}(x) h_{v_+}
(x)\} \big|0 \big\rangle\,\right]
\,,
\end{eqnarray}
where $v_+$ is the top four velocity ($v_+^2=1$) and $\hat s_t=2v_+.r$ 
and $h_{v_+}$ is the (HQET) heavy top quark field. The vacuum matrix
element also contains Wilson lines of the form
\begin{equation}
\label{bHQETWilsondef}
\nonumber
  W_n^\dagger(x) = {\rm P} \, 
   \exp\Big(i g\! \int_{0}^\infty \!\!\!\!ds\, \bn\cdot A_{+}(\bn s\!+\!x)
\Big) \,,\,\,\,
   W_n(x) =  \overline {\rm P} \:
   \exp\Big(- i g\! \int_{0}^\infty \!\!\!\!ds\, \bn\cdot
A_{+}(\bn s\!+\! x) \Big)\,,
\end{equation}
where $\bn$ is a light-like four vector pointing in the antitop
direction and $A_+$ is field describing a gluon that is collinear to
the quark.
Up to the Wilson lines the vacuum matrix element is in fact a heavy
quark propagator and, indeed, at tree-level it is just
\begin{equation}
\nonumber
  B^{\rm tree}_\pm(\hat s,\Gamma_t) 
  = {\rm Im} \bigg[\, \frac{-1}{\pi m_J}\: \frac{1}{\hat s-2\delta m+i\Gamma_t} \, \bigg]
  =   \frac{1}{\pi m_J} \: \frac{\Gamma_t}{\hat (s-2\delta m)^2 + \Gamma_t^2} \:,
\end{equation}
which is the imaginary part of the heavy quark propagator supplemented
by a constant width term and describing a Breit-Wigner distribution
having a width $\Gamma_t$. The residual mass term $\delta m$ becomes
relevant at higher orders and controls the mass scheme that is used.
For the pole mass scheme $\delta m=0$ to all order in $\alpha_s$.
It is the width term (which we can 
approximate as a constant since we are interested in the resonance
region) that allows us to use perturbation theory  for computing the
jet function. To understand the role of the Wilson lines
recall that the two-point function of simple heavy quark fields,
evaluated off-shell, is not gauge invariant, a fact that becomes
e.g.\,\,apparent from the gauge parameter dependence of the
perturbative corrections. The jet functions $B_\pm$, however, are
gauge-invariant due to the 
Wilson lines and well-defined {\it physical} objects. Physically the
Wilson lines describe gluons radiated from the antitop (moving along
the four vector $\bn$) that are collinear to the top quark (moving
along $v_+$), and it this additional radiation that renders the jet
function gauge-invariant and physically meaningful. In momentum space
the Wilson lines lead to additional Feynman diagrams having $1/(n\cdot
k\pm i 0)$ eikonal propagators.  

Having defined the jet function it is now straightforward to address
the question which mass scheme one might employ to have a good
perturbative behavior of the jet function. At
one-loop~\cite{Fleming:2007qr} one finds that the peak position is
located at  
$\hat s_{\rm peak}=2\delta
m-C_F\alpha_s(\mu)/2\Gamma_t[\ln(\frac{\mu}{\Gamma_t})+\frac{3}{2}]$. 
Recalling also that the pole mass contains a nasty ${\cal
  O}(\Lambda_{\rm QCD})$ renormalon, it therefore natural to use a mass
scheme different from the pole mass that is renormalon free and
absorbs at least the major part of the higher order corrections to the
peak position such that the resulting series is convergent. The
definition of such a scheme is obviously not unique and can also be
defined from moments of the distribution~\cite{inpreparatio}.
Generically we call such a mass a ``jet mass'' $m_J$ and its
perturbative relation to the pole mass reads
\begin{equation}
m_J = m^{\rm pole}_t - \delta m\,,
\end{equation}
where the HQET power counting requires that $\delta m\sim
\alpha_s \Gamma$ in the resonance region.
Using this relation one can relate the jet mass to other mass schemes.
The jet mass $m_J$ has already been used in the formulae shown before.
From this examination we see that top mass one can measure
from reconstruction is a jet mass. For sure, one cannot measure the
$\overline{\mbox{MS}}$ mass from reconstruction because it has $\delta m\sim
\alpha_s m_t\gg\Gamma_t$ and would invalidate the HQET power counting.

\section{Soft Function}

The soft function $S(\ell^+,\ell^-,\mu)$ describes the
non-perturbative contributions of the 
invariant mass distribution in the resonance region. Its definition
depends on the details of the prescription how soft radiation is
associated to $M_t$ and $M_{\bar t}$. 
 One possible prescription is
using a hemisphere mass definition, where $X_t$ and $X_{\bar t}$
contain everything to the left or right of the plane perpendicular to
the thrust axis of each event, see Fig.~\ref{fig:6topjet}. It is easy
to understand  that such a (and any other) prescription is leading to
a non-perturbative soft  function since one cannot compute
perturbatively how the soft particles are distributed around the
hemisphere boundary. Other prescriptions are possible as 
long as they do not associate soft radiation going in the top
direction to the antitop and vice-versa. This is in contrast to the jet
functions which, according to the condition on $X_{t}$ and $X_{\bar
  t}$ stated in Sec.~\ref{sectionfactheo}
are prescription-independent since they describe 
energetic jets within a small cone with opening angle $m_t/Q$ around
the top direction. At leading order in the power counting the
allowed prescriptions do not affect these energetic jets. For the
hemisphere prescription the soft function is defined by the vacuum
matrix element 
\begin{equation} 
\nonumber
  S_{\rm hemi}(\ell^+,\ell^-) = \frac{1}{N_c}\sum _{X_s} 
 \delta(\ell^+- k_s^{+a}) \delta(\ell^-- k_s^{-b})
  \langle 0| (\overline {Y}_\bn)^{cd}\,  ({Y}_n)^{ce} (0) |X_s \rangle
\langle X_s| ({Y}^\dagger_n)^{ef}\,  (\overline {Y}_\bn^\dagger)^{df}
(0) |0\rangle 
 \,,
\end{equation}
where $c,d,e,f$ are color indices and the $Y's$ are Wilson\
lines with soft gluons of the form 
\begin{eqnarray} \label{Yn}
 Y_n(x) =  \overline {\rm P} \:
   \exp\Big(- i g\! \int_{0}^\infty \!\!\!\!ds\, n\cdot A_{s}(ns\!+\! x) \Big) 
    \,,
 & Y_n^\dagger(x) =   {\rm P} \, 
   \exp\Big(\displaystyle i g\! \int_{0}^\infty \!\!\!\!ds\, n\cdot A_{s}(ns\!+\!x) \Big) \,,
  \nonumber\\
 \overline {Y_\bn}^\dagger(x)
  =   {\rm P} \: \exp\Big( i g\! \int_{0}^{\infty} \!\!\!\!ds\, 
      \bn\cdot \overline {A}_{s}(\bn s\!+\! x) \Big) 
  \,,  
  &\overline {Y_\bn}(x)  
   =   \overline {\rm P}\: \exp\Big(\displaystyle- i g\! \int_{0}^{\infty} \!\!\!\!ds\, 
      \bn\cdot \overline {A}_{s}(\bn s\!+\!x) \Big) 
   \,.
\end{eqnarray}
The $k_s^{+a}$ and $k_s^{-b}$ are operators that pick, according to
the hemisphere mass prescription, the total $+$
and $-$ light-cone momentum of the gluons that are in
hemisphere $a$ and $b$, respectively, see Fig.~\ref{fig:6topjet}.
These Wilson lines describe soft radiation off the top and antitop
quark and also render the soft function gauge-invariant. 

The factorization theorem (\ref{FactThm}) shows that the soft function
needs to be convoluted with the jet functions. This can be understood
physically, since the way how the soft radiation is associated to $M_t$
and $M_{\bar t}$ has to affect the observed invariant mass distribution. 
Field theoretically this convolution arises from the fact that the small
components of light-cone momenta in the top and antitop jets fluctuate
at the same length scales as the soft momenta described by the soft
function. At this point it is also useful to note that $S$ is a
renormalized object and that its renormalization group evolution can
be computed in perturbation theory. Nevertheless the actual form of
the soft function (i.e. the initial condition for the soft function
evolution at a low energy scale) is not computable with perturbative
methods. So in practice the soft function needs to be modeled and
eventually fixed by experimental data, similar to 
parton-distribution functions. How a soft model function can be
constructed incorporating consistently the required higher order
perturbative information has been discussed in Ref.~\cite{Hoang:2007vb}.

Given that the soft function is nonperturbative
and affects the invariant mass distribution at leading order, one
might ask what one has gained from predicting the invariant mass
distribution based on~(\ref{FactThm}) and concerning a precise
measurement of the top mass from the mass $M_{t,\bar t}$ where the
resonance is located. The crucial aspect is that the soft function is
universal and appears also in the factorization theorem for event
shape distributions for jets originating from massless quarks~\cite{Gehrmann}  
in the dijet region, where the thrust $T\approx
1$~\cite{Korchemsky:1994is}. This is related to
the fact that the soft $Y$ Wilson lines that arise from massless and from
massive quark lines are identical.
So our factorization theorem for the top invariant mass distribution
in the resonance region becomes predictive after having determined a
soft function from event  shape distributions from $e^+e^-$ 
data, which are already available from LEP~\cite{Kluth:2006bw}. Such
a determination of the soft function was carried out by Korchemsky and
Tafat in Ref.~\cite{Korchemsky:2000kp}.

\section{Numerical Analysis at LO}

Using the factorization theorem it is straightforward to carry out a
simple LO analysis using the tree-level result for the jet functions
(i.e. one can set $\delta m=0$) and the soft model function determined
by Korchemsky and Tafat:
\begin{equation} \label{SM1}
  S_{\rm hemi}^{\rm M1}(\ell^+,\ell^-) = \theta(\ell^+)\theta(\ell^-)
   \frac{ {\cal N}(a,b) }{\Lambda^2}
  \Big( \frac{\ell^+\ell^-}{\Lambda^2}\Big)^{a-1} \exp\Big(
  \frac{-(\ell^+)^2-(\ell^-)^2-2 b \ell^+\ell^-}{\Lambda^2} \Big)\,,
\end{equation}
where ${\cal N}$ is a normalization factor. From fits to $e^+e^-$
heavy jet mass and thrust LEP data they obtained 
\begin{equation}\label{abL}
 a=2 \,, \qquad\quad
 b=-0.4 \,, \qquad\quad
 \Lambda=0.55\,{\rm GeV} \,,
\end{equation}
which we adopt in the following. The analysis illustrates a number
of important features related to how the predictions by
the factorization theorem depend on the c.m.\,energy Q.

\begin{figure}[t!]
% a=2, b=-0.4
  \centerline{ 
   \includegraphics[width=6cm]{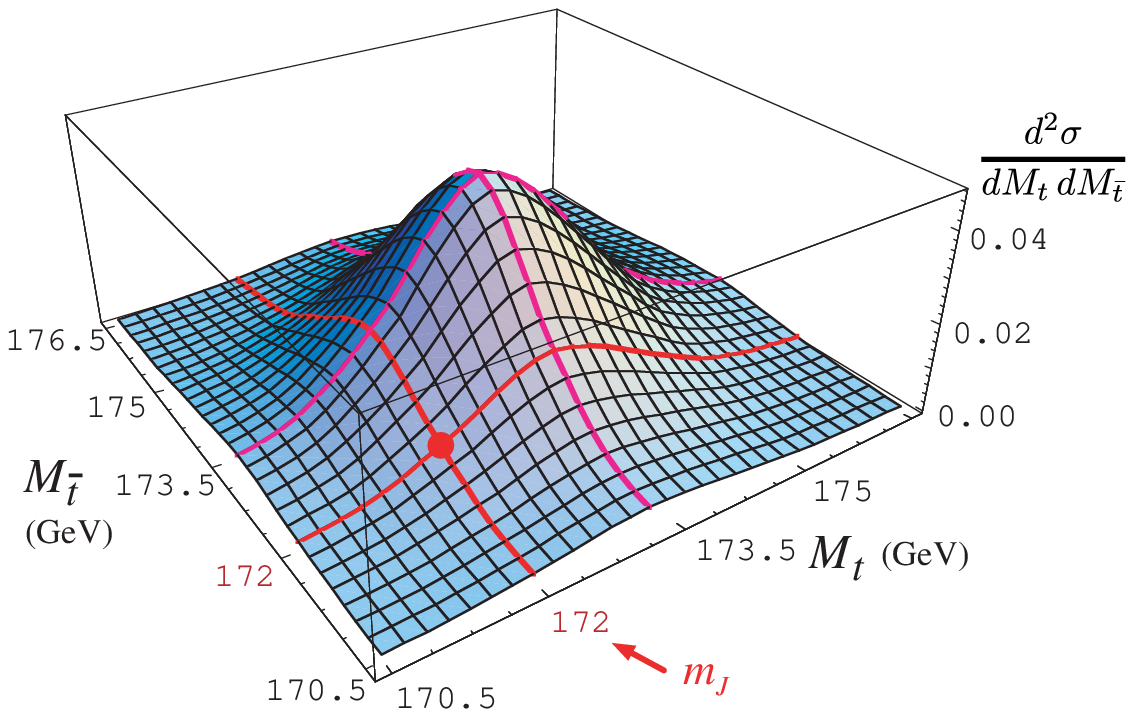}
   \qquad\qquad
   \includegraphics[width=4.6cm]{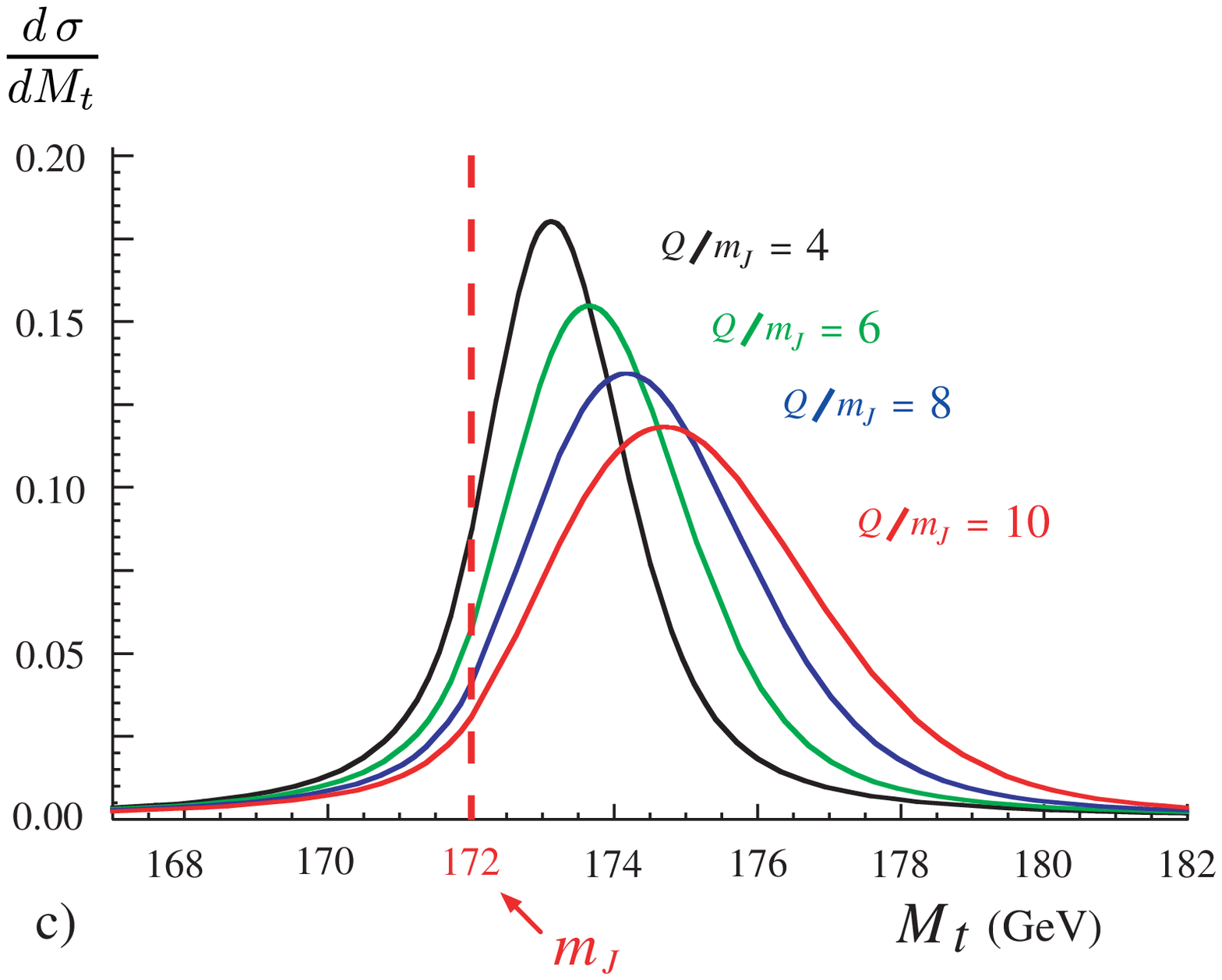}
  } 
\caption{(a) Plot of the double differential hemisphere invariant mass
  cross-section $d^2\sigma/dM_t dM_{\bar t}$ in units of
  $4\sigma_0/\Gamma_t^2$ for $m_J=172$, $Q=4.33
  m_J$ and $\Gamma_t=1.43$~GeV.
(b) Dependence of the single differential invariant mass distribution as
described in the text on the c.m.\,\,energy $Q$ with the same
normalization.}
\label{fig:plot3D}
\end{figure}

In Fig.~\ref{fig:plot3D}a the double differential invariant mass
distribution is displayed for the input values $m_J=172$, $Q=4.33
  m_J$ and $\Gamma_t=1.43$~GeV.
The conspicuous feature of the predicted distribution is that the
observable resonance peak it shifted toward a higher value by about
$1.5$~GeV. This feature is one of the important properties of a
invariant mass prescription that assigns all soft radiation to the
masses. In the factorization theorem it arises from the $Q/m$ factor
involved in the convolution over the variables $\ell^\pm$. Intuitively
it can be easily understood from the fact that the total
invariant mass of a fast moving particle with mass $m$ plus a soft
momentum increases linearly with the soft momentum and the boost
factor of the massive particle. This feature is also visible in
Fig.~\ref{fig:plot3D}b where the single differential invariant mass
distribution 
\begin{equation}
\frac{d\sigma}{d M_t} = \frac{2}{\Gamma} \int_{M_{\rm lower}}^{M_{\rm upper}} dM_{\bar t}\
  \:    \frac{d^2\sigma}{d M_t d M_{\bar t}},
\end{equation}
is plotted. Here the integration interval $[M_{\rm lower},M_{\rm upper}]$ is
twice the size of the measured peak mean half width and centered at the peak
mass. For the single differential distribution one can relate the peak
location approximately to the first moment of the soft function by
\begin{equation}
 M_t^{\rm peak}\simeq m_J + \frac{Q}{2m_J}\, S_{\rm hemi}^{[1,0]}.
\end{equation}
Interestingly this relation can also be used for fixed $Q/m$ for invariant
mass prescriptions that differ in the treatment of the soft radiation and
lead to different first moments of the soft function. When extrapolated to
zero moment linearly one can obtain an estimate for the jet mass.

We also find that the distribution gets wider with 
$Q/m$. This is again a consequence of the $Q/m$ factor occurring in the
convolution over $\ell^\pm$ in the factorization theorem since 
for increasing $Q$ the jet function gets effectively smeared over
a wider distribution. The shift of the peak position as well as the widening
of the invariant mass distribution have been observed in simulation studies 
at the ILC~\cite{Chekanov:2002sa} and the LHC for large $p_T$
events~\cite{Borjanovic:2004ce} and can now be better quantified using the 
factorization theorem.

\section{Acknowledgments}
This work was supported in part by the EU network contract
  MRTN-CT-2006-035482 (FLAVIAnet).

% ****************************************************************************
% BIBLIOGRAPHY AREA
% ****************************************************************************

\begin{footnotesize}
% IF YOU DO NOT USE BIBTEX, USE THE FOLLOWING SAMPLE SCHEME FOR THE REFERENCES
% ----------------------------------------------------------------------------

% ----------------------------------------------------------------------------

% IF YOU USE BIBTEX,
% - DELETE THE TEXT BETWEEN THE TWO ABOVE DASHED LINES
% - UNCOMMENT THE NEXT TWO LINES AND REPLACE 'Name_Of_Your_BibFile'

%\bibliographystyle{unsrt}
%\bibliography{Name_Of_Your_BibFile}

\begin{thebibliography}{99}
% Please replace the numbers for   contribId   and   sessionId
% in the following URL. You can get this information by going to 
% http://indico.cern.ch/confAuthorIndex.py?confId=9499
% and search for your contribution and click on the title
% Be aware: '&amp;' must be replaced by simple '&' as in example below
\bibitem{url} Slides: \\ 
\verb$http://ilcagenda.linearcollider.org/contributionDisplay.py?contribId=190&sessionId=85&confId=1296$
\\
\verb$http://ilcagenda.linearcollider.org/contributionDisplay.py?contribId=408&sessionId=73&confId=1296$


%------- replace following references ;-)
\bibitem{Hoang:2000yr}
  A.~H.~Hoang {\it et al.},
  %``Top-antitop pair production close to threshold: Synopsis of recent NNLO
  %results,''
  Eur.\ Phys.\ J.\ direct C {\bf 2}, 1 (2000)
  [arXiv:hep-ph/0001286];
  %%CITATION = EPHJD,C2,1;%%
%\bibitem{Hoang:2003xg}
  A.~H.~Hoang,
  %``Top Pair Production at Threshold and Effective Theories,''
  Acta Phys.\ Polon.\  B {\bf 34}, 4491 (2003)
  [arXiv:hep-ph/0310301].
  %%CITATION = APPOA,B34,4491;%%
%
\bibitem{Kiyo} Y.~Kiyo, these proceedings.
%
\bibitem{AHH} A.~H. Hoang, these proceedings.
%
\bibitem{Gounaris} F.~Gounaris, these proceedings.
%
\bibitem{Boogert} S.~Boogert, these proceedings.
%
\bibitem{Chekanov:2002sa}
  S.~V.~Chekanov,
  %``Uncertainties on the measurements of the top mass at a future e+ e-
  %collider,''
  arXiv:hep-ph/0206264;
  %%CITATION = HEP-PH/0206264;%%
%bibitem{Chekanov:2003cp}
  S.~V.~Chekanov and V.~L.~Morgunov,
  %``Selection and reconstruction of the top quarks in the all-hadronic  decays
  %at a linear collider,''
  Phys.\ Rev.\  D {\bf 67}, 074011 (2003)
  [arXiv:hep-ex/0301014].
  %%CITATION = PHRVA,D67,074011;%%
%
\bibitem{Fleming:2007qr}
  S.~Fleming, A.~H.~Hoang, S.~Mantry and I.~W.~Stewart,
  %``Jets from Massive Unstable Particles: Top-Mass Determination,''
  arXiv:hep-ph/0703207.
  %%CITATION = HEP-PH/0703207;%%
%
\bibitem{Borjanovic:2004ce}
  I.~Borjanovic {\it et al.},
  %``Investigation of top mass measurements with the ATLAS detector at LHC,''
  Eur.\ Phys.\ J.\  C {\bf 39S2}, 63 (2005)
  [arXiv:hep-ex/0403021].
  %%CITATION = EPHJA,C39S2,63;%%
%
\bibitem{Bauer:2000yr}
  C.~W.~Bauer, S.~Fleming and M.~E.~Luke,
  %``Summing Sudakov logarithms in B --> X/s gamma in effective field  theory,''
  Phys.\ Rev.\  D {\bf 63}, 014006 (2001)
  [arXiv:hep-ph/0005275];
  %%CITATION = PHRVA,D63,014006;%%
  C.~W.~Bauer, S.~Fleming, D.~Pirjol and I.~W.~Stewart,
  %``An effective field theory for collinear and soft gluons: Heavy to light
  %decays,''
  Phys.\ Rev.\  D {\bf 63}, 114020 (2001)
  [arXiv:hep-ph/0011336];
  %%CITATION = PHRVA,D63,114020;%%
  C.~W.~Bauer, D.~Pirjol and I.~W.~Stewart,
  %``Soft-collinear factorization in effective field theory,''
  Phys.\ Rev.\  D {\bf 65}, 054022 (2002)
  [arXiv:hep-ph/0109045].
  %%CITATION = PHRVA,D65,054022;%%
  %\bibitem{Bauer:2002nz}
  C.~W.~Bauer, S.~Fleming, D.~Pirjol, I.~Z.~Rothstein and I.~W.~Stewart,
  %``Hard scattering factorization from effective field theory,''
  Phys.\ Rev.\  D {\bf 66}, 014017 (2002)
  [arXiv:hep-ph/0202088].
  %%CITATION = PHRVA,D66,014017;%%
%
\bibitem{Manohar:2000dt}
  A.~V.~Manohar and M.~B.~Wise,
  %``Heavy quark physics,''
  Camb.\ Monogr.\ Part.\ Phys.\ Nucl.\ Phys.\ Cosmol.\  {\bf 10} (2000) 1.
  %%CITATION = CMPCE,10,1;%%
%
\bibitem{Fadin:1988fn}
  V.~S.~Fadin and V.~A.~Khoze,
  %``Production of a pair of heavy quarks in e+ e- annihilation in the threshold
  %region,''
  Sov.\ J.\ Nucl.\ Phys.\  {\bf 48}, 309 (1988)
  [Yad.\ Fiz.\  {\bf 48}, 487 (1988)];
  %%CITATION = YAFIA,48,487;%%
  M.~Beneke, A.~P.~Chapovsky, A.~Signer and G.~Zanderighi,
  %``Effective theory approach to unstable particle production,''
  Phys.\ Rev.\ Lett.\  {\bf 93}, 011602 (2004)
  [arXiv:hep-ph/0312331];
  %%CITATION = PRLTA,93,011602;%%
   A.~H.~Hoang and C.~J.~Reisser,
  %``Electroweak absorptive parts in NRQCD matching conditions,''
  Phys.\ Rev.\  D {\bf 71}, 074022 (2005)
  [arXiv:hep-ph/0412258].
  %%CITATION = PHRVA,D71,074022;%%
%
\bibitem{inpreparatio}
 S.~Fleming, A.~H.~Hoang, S.~Mantry and I.~W.~Stewart, in preparation
 (2007)
%
\bibitem{Hoang:2007vb}
  A.~H.~Hoang and I.~W.~Stewart,
  %``Designing Gapped Soft Functions for Jet Production,''
  arXiv:0709.3519 [hep-ph].
  %%CITATION = ARXIV:0709.3519;%%
%
\bibitem{Gehrmann} T.~Gehrmann, these proceedings.
%
\bibitem{Korchemsky:1994is}
  G.~P.~Korchemsky and G.~Sterman,
  %``Nonperturbative corrections in resummed cross-sections,''
  Nucl.\ Phys.\  B {\bf 437}, 415 (1995)
  [arXiv:hep-ph/9411211];
  %%CITATION = NUPHA,B437,415;%%
%\bibitem{Bauer:2003di}
  C.~W.~Bauer, C.~Lee, A.~V.~Manohar and M.~B.~Wise,
  %``Enhanced nonperturbative effects in Z decays to hadrons,''
  Phys.\ Rev.\  D {\bf 70}, 034014 (2004)
  [arXiv:hep-ph/0309278].
  %%CITATION = PHRVA,D70,034014;%%
\bibitem{Kluth:2006bw}
  S.~Kluth,
  %``Tests of quantum chromo dynamics at e+ e- colliders,''
  Rept.\ Prog.\ Phys.\  {\bf 69}, 1771 (2006)
  [arXiv:hep-ex/0603011].
  %%CITATION = RPPHA,69,1771;%%
%
\bibitem{Korchemsky:2000kp}
  G.~P.~Korchemsky and S.~Tafat,
  %``On power corrections to the event shape distributions in QCD,''
  JHEP {\bf 0010}, 010 (2000)
  [arXiv:hep-ph/0007005];
  %%CITATION = JHEPA,0010,010;%%




\end{thebibliography}
% example of Name_Of_Your_BibFile.bib
% @Article{Turcato:2006ch,
%      author    = "Turcato, M.",
%  collaboration = "ZEUS and H1",
%      title     = "Lepton flavour violation and charmonium physics at HERA",
%      journal   = "Nucl. Phys. Proc. Suppl.",
%      volume    = "162",
%      year      = "2006", 
%      pages     = "283-287",
%      SLACcitation  = "%%CITATION = NUPHZ,162,283;%%"
% }
% 
% @Unpublished{Gogitidze:2007du,
%      author    = "Gogitidze, N.",
%  collaboration = "H1", 
%      title     = "Prompt photons and particle momentum distributions at
%                   HERA", 
%      year      = "2007",
%      note    = "hep-ex/0701033",
%      SLACcitation  = "%%CITATION = HEP-EX 0701033;%%"
% }

\end{footnotesize}

% ****************************************************************************
% END OF BIBLIOGRAPHY AREA
% ****************************************************************************

\end{document}